\documentclass[journal]{IEEEtran}  

\IEEEoverridecommandlockouts                              

\usepackage{tikz}
\usetikzlibrary{shapes, arrows.meta, positioning, calc}
\usepackage{pgfplots}
\usepackage{graphicx} 
\usepackage{setspace}
\usepackage{xcolor}
\usepackage{amsmath}
\usepackage{amsfonts}
\usepackage{bbm}
\usepackage{mathtools} 
\RequirePackage{algorithm}
\RequirePackage{algpseudocode} 
\usepackage{caption}
\usepackage{subcaption}
\usepackage{comment}
\usepackage{listings}
\usepackage{dsfont}
\usepackage{amsfonts}
\usepackage{booktabs}
\usepackage{optidef}
\usepackage{hyperref}
\usepackage{url}
\usepackage{bm}

\usepackage{amsthm}

\theoremstyle{definition}    

\newcommand{\R}[1]{\mathrm{#1}}		
	
\newcommand{\T}[2]{{\mathbf{T}_{#1}}(#2)}
\newcommand{\dT}[2]{{\dot{\mathbf{T}}_{#1}}(#2)}
\newcommand{\ddT}[2]{{\ddot{\mathbf{T}}_{#1}}(#2)}
\newcommand{\He}[2]{{\mathbf{H}_{#1}}(#2)}
\newcommand{\dHe}[2]{{\dot{\mathbf{H}}_{#1}}(#2)}

\newcommand{\numherders}[0]{N}
\newcommand{\numtargets}[0]{M}

\newcommand{\goalcenter}[1]{\mathbf{c}_{\mathrm{G}}(#1)}
\newcommand{\goalradius}[0]{\rho_{\mathrm{G}}}
\newcommand{\domainradius}[0]{\rho_{\mathrm{0}}}
\newcommand{\domain}[0]{\Omega}
\newcommand{\initialdomain}[0]{\domain_\mathrm{0}}
\newcommand{\goalregion}[0]{\domain_{\mathrm{G}}}

\newcommand{\shortrangerep}[2]{\boldsymbol{\varphi}_{#1}(#2)}
\newcommand{\longrangerep}[2]{\boldsymbol{\gamma}_{#1}(#2)}
\newcommand{\repulsionpotential}[0]{\boldsymbol{\Gamma}}

\newcommand{\neighbouragents}[0]{\Phi}

\newcommand{\diffusioncoeff}[0]{\sigma}
\newcommand{\shortrangecoeff}[0]{\alpha}
\newcommand{\longrangecoeff}[0]{\beta}

\newcommand{\herdermaxvel}[0]{v_\mathrm{H}}

\newcommand{\samplingtime}[0]{\Delta t}
\newcommand{\samplingtimetrain}[0]{\Delta \Tilde{t}}

\newcommand{\finaltime}[0]{t_{\mathrm{f}}}
\newcommand{\settlingtime}[0]{t^{\star}}
\newcommand{\gatheringtime}[0]{t_\R{g}}
\newcommand{\containmenttime}[0]{\Delta t_\R{c}}
\newcommand{\maxtime}[0]{t_\R{h}}
\newcommand{\pathlength}[1]{d(#1)}
\newcommand{\pathlengthfinal}[0]{d_\mathrm{f}}
\newcommand{\pathlengthgath}[0]{d_\mathrm{g}}
\newcommand{\coopratio}[1]{\Psi(#1)}
\newcommand{\coopratiofinal}[0]{\Psi_\mathrm{f}}
\newcommand{\coopratiogath}[0]{\Psi_\mathrm{g}}

\newcommand{\numepisodes}[0]{E}

\newcommand{\selectedtarget}[2]{\mathbf{T}^*_{#1}(#2)}

\usepackage{dsfont}
\newcommand{\indicator}[2]{\mathds{1}_{#1}\left(#2\right)}
\newcommand{\B}[1]{\if#1\relax\bm{#1}\else\mathbf{#1}\fi} 

\newcommand{\norm}[1]{\left\lVert#1\right\rVert}

\definecolor{myorange}{RGB}{205, 102, 57}
\pgfplotsset{compat=1.18} 

\usepackage{csquotes}

\title{\LARGE \bf
Hierarchical Learning-Based  Control for Multi-Agent Shepherding of Stochastic Autonomous Agents%
}

\author{Italo Napolitano\textsuperscript{1,\dag}, Stefano Covone\textsuperscript{1,\dag}, Andrea Lama\textsuperscript{1}, Francesco De Lellis\textsuperscript{2}, Mario di Bernardo\textsuperscript{1,2,*} 
\thanks{\textsuperscript{1}Scuola Superiore Meridionale, Naples, Italy}%
\thanks{\textsuperscript{2}Department of Electrical Engineering and Information Technology, University of Naples Federico II, Naples, Italy}%
\thanks{\textsuperscript{\dag} These authors contributed equally to this work. }
\thanks{\textsuperscript{*} Corresponding author}}

\begin{document}

\maketitle
\thispagestyle{empty}
\pagestyle{empty}


\begin{abstract}
Multi-agent shepherding represents a challenging distributed control problem where herder agents must coordinate to guide independently moving targets to desired spatial configurations. Most existing control strategies assume cohesive target behavior, which frequently fails in practical applications where targets exhibit stochastic autonomous behavior. This paper presents a hierarchical learning-based control architecture that decomposes the shepherding problem into a high-level decision-making module and a low-level motion control component. The proposed distributed control system synthesizes effective control policies directly from closed-loop experience without requiring explicit inter-agent communication or prior knowledge of target dynamics. The decentralized architecture achieves cooperative control behavior through emergent coordination without centralized supervision. Experimental validation demonstrates superior closed-loop performance compared to state-of-the-art heuristic control methods, achieving 100\% success rates with improved settling times and control efficiency. The control architecture scales beyond its design conditions, adapts to time-varying goal regions, and demonstrates practical implementation feasibility through real-time experiments on the Robotarium platform.
\end{abstract}

\section{Introduction}
\label{sec:introduction}
\IEEEPARstart{T}{he} 
control of large groups of autonomous agents represents one of the most challenging problems in modern distributed control systems. Multi-agent coordination finds critical applications in autonomous vehicle fleets, industrial automation, and emergency response operations, where centralized control approaches become computationally intractable or communication-limited.

A paradigmatic example  of this challenge is the multi-agent {\em shepherding control problem}, in which a team of agents, or \textit{herders}, must steer the collective dynamics of another group of agents, or \textit{targets}, toward desired spatial configurations \cite{long2020comprehensive}. 
Bio-inspired by the way shepherd dogs guide sheep, this paradigm finds relevant applications in mine sweeping, area defense, museum guidance, oil-spill containment \cite{lien2005shepherding,chipadeAerialSwarmDefense2021,zahugi2013oil} and disaster response where autonomous herders could steer livestock away from flood zones \cite{paranjape2018robotic}.

 Unlike traditional formation control or consensus problems where all agents cooperate toward common objectives, shepherding involves non-cooperative targets moving according to autonomous dynamics.  
 This creates unique challenges: (i) heterogeneous agent populations with different dynamics,  
 (ii) indirect influence through environmental interactions \cite{licitraSingleAgentIndirectherding2018}, (iii) complex emergent behaviors difficult to model analytically, and (iv) real-time distributed decision-making under communication constraints.
 
Current shepherding strategies rely on heuristic coordination rules or model-based predictive control \cite{long2020comprehensive} often inspired by animal decision-making processes \cite{haque2011biologically}. 
Heuristic methods lack optimality and may perform poorly under varying conditions. Model-based strategies require detailed knowledge of target dynamics, which may be unavailable or time-varying. Furthermore, designing model-based strategies without predefined maneuvers remains challenging or limited to relatively small number of agents \cite{piersonControllingNoncooperativeHerds2018}.

Furthermore, most existing approaches assume that targets exhibit cohesive collective behavior \cite{strombomSolvingShepherdingProblem2014, van2024reactive}, enabling the group to be treated as a single controllable entity. 
Common strategies involve collecting targets into clusters, then driving the group toward goals  \cite{strombomSolvingShepherdingProblem2014}. This approach has influenced several subsequent studies \cite{zheng2024bio, li2023communication}, whereas formation control methods adopt a different strategy for guiding the herd \cite{piersonControllingNoncooperativeHerds2018}.

However, cohesive behavior frequently fails in panic situations, wildlife management, or heterogeneous robotic swarms \cite{zhang2024distributed}, \cite{lamaShepherdingControlHerdability2024}. Without cohesion, herders must influence each target individually, dramatically increasing control complexity, as noted in \cite{koAsymptoticBehaviorControl2020}. 

Only few heuristic solutions address non-cohesive targets \cite{lamaShepherdingControlHerdability2024,aulettaHerdingStochasticAutonomous2022} creating need for novel architectures that learn effective coordination without detailed system models or restrictive assumptions.

More recently, learning-based frameworks, particularly Reinforcement Learning, have emerged to approximate optimal control strategies beyond rule-based heuristics. Most RL approaches to shepherding  assume the presence of a single herder and cohesive targets, and employ heuristic behaviors \cite{nguyen2019deep, hussein2022autonomous}, while multi-herder extensions use multi-task reinforcement learning \cite{wang2024multi}. Deep Q-Networks (DQN) have been trained on surrogate potential fields for cohesive herds \cite{zhi2021learning}, while decentralized Multi-Agent Reinforcement Learning (MARL) approaches use Proximal Policy Optimization (PPO) in Centralized-Training-Decentralized-Execution (CTDE) frameworks for payload protection scenarios \cite{hasanFlockNavigation2023}. 

Only a few studies have explored machine learning-based control strategies that relax the cohesion assumption. Some of these approaches train herders in single-target scenarios and then extend the learned behaviors to multi-target settings using heuristic rules \cite{ninoDeepAdaptiveIndirect2023, delellisApplicationControlTutored2021}, though such methods often yield suboptimal performance. Within proper MARL frameworks, recent work uses Dynamical Perceptual-Motor Primitives with PPO for target selection \cite{patil2023scaffolding}, but severely constrains herder behavior and assumes deterministic targets.

To address these control system design challenges and modeling limitations, this paper presents a hierarchical learning-based control architecture for multi-agent shepherding of stochastic (non-cohesive) autonomous agents. 
The proposed approach breaks down the complex coordination task into manageable layers: a high-level decision-making module that assigns targets to herders, and a low-level controller that computes the herders’ movements to guide targets toward the desired goal region.
Each layer employs reinforcement learning to synthesize control policies directly from closed-loop experience, eliminating requirements for explicit system models or inter-agent communication.

This manuscript significantly extends our preliminary results in \cite{covone2025hierarchical,napolitano2024emergentcooperativestrategiesmultiagent} by presenting the first complete, validated model-free framework for multi-agent shepherding. While existing methods have addressed either emergent cooperation through DQN or hierarchical policies via PPO independently, this work integrates both approaches into a unified system with comprehensive real-world validation.
In particular, the key contributions of this paper include: (i) developing a complete hierarchical framework to solve the shepherding problem without cohesion assumptions, with systematic comparison of DQN versus PPO across both driving and target selection policies, (ii) comprehensive technical validation through benchmarking against state-of-the-art approaches and extensive robustness analysis demonstrating superior performance, (iii) extending the framework's applicability through scalability to large-scale scenarios with limited sensing and adaptation to time-varying goal regions, and (iv) thorough experimental validation on the Robotarium platform demonstrating real-world feasibility. Herders' cooperative behavior is shown to emerge naturally from the learning process without explicit coordination, demonstrating practical applicability for real-world multi-robot systems.

\section{Problem statement}
\label{sec: mathematization}
We consider a spatial domain $\domain \subseteq \mathbb{R}^2$ populated by two interacting agents' populations:
$\numherders$ controlled \textit{herders}, whose task is to guide
$\numtargets$  \textit{targets} toward a goal region $\goalregion \subset \domain$.

Let
\(
\He{}{t} = \left[ \He{1}{t}, \dots, \He{\numherders}{t} \right] \in \domain^\numherders,
\) denote the vector of herders' positions at time $t$,
where $\He{j}{t} \in \domain$ represent the Cartesian coordinates of the $j$-th herder. Similarly, target positions are denoted by $\T{}{t} = \left[ \T{1}{t}, \dots, \T{\numtargets}{t} \right] \in \domain^\numtargets$, with $\T{i}{t} \in \domain$ being the position of the $i$-th target agent at time $t$. We refer to a generic agent $\mathbf{X}_a(t) \in \{\T{a}{t}, \He{a}{t}\}$ and define $\mathbf{X}(t) = \left[\T{}{t}, \He{}{t}\right] \in \domain^{\numherders+\numtargets}$.

Following \cite{lamaShepherdingControlHerdability2024}, we consider an unbounded domain $\domain \equiv \mathbb{R}^2$ with agents initially placed uniformly at random within a circular region $\initialdomain \subset \domain$ of radius $\domainradius \in \mathbb{R}^+$. The goal region $\goalregion$ is defined as a disk of radius $\goalradius < \domainradius$, centered at $\goalcenter{t} \in \domain$. Without loss of generality, we consider a static goal region centered at the origin, $\goalcenter{t} = \mathbf{0}_2$, which is later generalized to time-varying goal regions.

\paragraph{Assumptions}
To formulate the problem, we make the following modeling assumptions:
\begin{enumerate}
    \item All agents exhibit short-range repulsion to prevent collisions;

    \item Targets follow second-order stochastic dynamics \cite{albi2016} and do not exhibit cohesive collective behavior (e.g., flocking).
    \item Herders exert long-range repulsive forces on targets and follow first-order dynamics with bounded control inputs. This is standard in shepherding \cite{piersonControllingNoncooperativeHerds2018} where inertial effects are typically neglected \cite{albi2016}.
    \item Herders have access to the positions of all agents and the center of the goal region.
    \item Agents within the same population are homogeneous.
\end{enumerate}

\paragraph{Targets' Dynamics}
Under the assumptions above, the targets follow the Langevin equation:
\begin{equation}
\begin{aligned}
    \ddT{i}{t} = & -\zeta \dT{i}{t} 
    + \shortrangerep{}{\T{i}{t}, \mathbf{X}(t)} + \\
    & + \longrangerep{}{\T{i}{t}, \He{}{t}}
    + \diffusioncoeff\bm{\mathcal{N}}_i(t),
    \label{eqn: target_dyn}
\end{aligned}
\end{equation}
where $\zeta>0$ is the damping coefficient, $\bm{\mathcal{N}}(t)$ is Gaussian noise with unitary variance, and $\diffusioncoeff>0$ regulates the noise strength.

As typically done in the literature, see e.g. \cite{aulettaHerdingStochasticAutonomous2022,piersonControllingNoncooperativeHerds2018,sebastian_adaptive_2022,zhang2024distributed}, we define the herder-target repulsion from a potential field
\begin{equation}
    \label{eqn:rep_potential}
    \repulsionpotential(\T{}{t}, \He{}{t}) = -\sum_{i=1}^{\numtargets} \sum_{j=1}^{\numherders}  \frac{1}{|| \T{i}{t} - \He{j}{t} ||},
\end{equation}
yielding the repulsive force
\begin{equation}
\begin{aligned}
\label{eqn:long_range_repulsion}
    \longrangerep{}{\T{i}{t}, \He{}{t}} &= \longrangecoeff \frac{\partial}{\partial \T{i}{t}} \Gamma (\T{}{t}, \He{}{t})    = \\
    &= - \longrangecoeff \sum_{j=1}^\numherders 
    \frac{\T{i}{t} - \He{j}{t}}{\norm{\T{i}{t} - \He{j}{t}}^3},
\end{aligned}
\end{equation}
where $\longrangecoeff > 0$ is the interaction strength. Variants of the model may adopt different expressions for the potential $\repulsionpotential$ as, for example, in \cite{van2024reactive, lamaShepherdingControlHerdability2024, koAsymptoticBehaviorControl2020}. 

Similarly, as in \cite{aulettaHerdingStochasticAutonomous2022, zhang2024distributed}, every agent $\mathbf{X}_i$ experiences short-range repulsive forces from neighboring agents $\{\mathbf{X}_j\}_{j \neq i}$:
\begin{equation}
     \shortrangerep{}{\B{X}_i(t), \B{X}(t)} = \shortrangecoeff \sum_{j \in \neighbouragents_i}  \frac{ \B{X}_i(t) - \B{X}_j(t) }{\norm{\B{X}_i(t) - \B{X}_j(t)}^3},
     \label{eqn:short_range_repulsion}
\end{equation}
where $\shortrangecoeff > 0$ is the interaction strength and
$\neighbouragents_i (t) = \left\{  j \neq i: \norm{ \B{X}_j (t) - \B{X}_i (t) } \leq  r_c \right\}$ is the set of agents closer than a safety radius $r_c$. 

\paragraph{Herders' Dynamics}
Herders are modeled as single integrators, as commonly done in robotic control \cite{piersonControllingNoncooperativeHerds2018}:
\begin{equation}
    \dHe{j}{t} =  
    \shortrangerep{}{\He{j}{t}, \B{X}(t)} + 
    \mathbf{u}_j(t),
    \label{eqn: herder_dyn}
\end{equation}
where $\mathbf{u}_j \in \left[ - \herdermaxvel,  \herdermaxvel \right]^2$ is the control action, while $\shortrangerep{}{\cdot}$ is the short-range repulsion in \eqref{eqn:short_range_repulsion}.

Parameter values for both herders' and targets' models are reported in Appendix \ref{appendix:numerical}.

\paragraph{Control Objective}
The control objective is to design a decentralized control law for the herders, such that all targets are driven and contained within the goal region. Herders rely solely on observations about other agents' positions, without communicating internal decisions, consistent with the assumptions in, e.g., \cite{li2023communication}. Moreover, we assume unknown models of the agents when designing control policies.

We formalize the control goal as follows.
Following \cite{lamaShepherdingControlHerdability2024}, let $\chi(t)$ define the fraction of targets inside the goal region at time $t$:
\begin{equation}
    \chi(t) = \frac{ \mid \{ i : \mathbf{T}_i(t) \in  \goalregion, \; i \in \left[1, M\right]\} \mid}{\numtargets},
    \label{eq: chi}
\end{equation}
where $\mid \mathcal{A} \mid$ denotes the cardinality of set $\mathcal{A}$.

We aim to design a policy $\pi(\mathbf{u}_j \mid \mathbf{S}_j)$ for generating control actions $\mathbf{u}_j \sim \pi(\cdot \mid \mathbf{S}_j) \; \forall j = 1, \dots, \numherders$ such that
\begin{equation}
    \exists \, \bar t < \infty \text{ s.t. } \chi(\tau) \geq \chi^* \; \forall \tau \geq \bar t,
\end{equation}
where (i) $\mathbf{S}_j \in \domain^{\numherders + \numtargets}$ denotes the $j$-th herder’s observation vector, comprising the positions of its sensed agents, and (ii) $\chi^*$ denotes the desired minimum fraction of targets within the goal region (e.g., $\chi^*=0.99$).

To align with the discrete-time nature of controllers and actuators, we reformulate the control problem in discrete time, where time instants are denoted as $t_k=k \Delta t$, with $k$ being the decision step and $\Delta t$ the sampling interval.

\subsection{Metrics}
\label{sec: metrics}
To evaluate how effectively a candidate policy satisfies the above objective, we introduce the following metrics, where we consider a value of $\chi^* = 0.99$:
\begin{itemize}
    \item {\em Gathering time} $\gatheringtime$. First time instant in which all targets enter $\goalregion$:
    \begin{equation}
        \gatheringtime = \inf_{t} \{ t \geq 0  \; \text{:} \; \chi(t) \geq \chi^* \},
    \end{equation}
    
    \item {\em Settling time} $\settlingtime$. First time instant when all targets enter and subsequently remain in $\goalregion$:
    \begin{equation}
        \settlingtime = \inf_{t} \{ t \geq 0  \; \text{:} \; \chi(t_k) \geq \chi^*, \forall t_{k} \in \left[ t, \finaltime \right] \},
    \end{equation}
    where $\finaltime = \min \left( t + \containmenttime, \maxtime \right)$. An episode terminates when all targets remain within the goal region for a time $\containmenttime$ or if the maximum time $\maxtime$ is reached.
    The problem is solved if $\settlingtime$ is finite.
    
    \item {\em Average Path Length} $\pathlength{t}$. Mean distance travelled by each herder in the interval $\left[0, t \right]$.
    \begin{equation}
    \label{eq:path_length}
        \pathlength{t} = \frac{1}{\numherders} \sum_{i=1}^\numherders \int_{0}^{t} \norm{\dHe{i}{\tau}} \mathrm{d}\tau.
    \end{equation}
    Note that this also serves as a good proxy for the average control effort, since $\dHe{}{t} \approx \mathbf{u}(t)$ (cf. Eq. \eqref{eqn: herder_dyn}), with only a negligible contribution from collision avoidance.
    In our simulations, we evaluate  $\pathlengthfinal = d(\finaltime)$ and $\pathlengthgath = \pathlength{\gatheringtime}$. 
    
    \item {\em Average cooperation index} $\coopratio{t}$, which quantifies the degree of cooperation among herders in pursuing distinct targets. 
     It is defined as the average over the time interval $[0,\ t]$, of the ratio between the number of different pursued targets and the number of herders:
    \begin{equation}
        \coopratio{t} = \frac{1}{t} \int_{0}^{t} \frac{|\mathcal{S}(\tau)| - 1}{N - 1} \mathrm{d}\tau,
    \end{equation}    
    where $\mathcal{S}(\tau)$ is the set of different pursued targets at time $\tau$.
    If $\psi \sim 1$, the herders cooperate by almost always selecting different targets, whereas a $\psi \sim 0$ indicates that the herders tend to select the same target.
    In particular, we evaluate $\coopratiofinal = \coopratio{\finaltime}$ and $\coopratiogath = \coopratio{\gatheringtime}$. 
\end{itemize}

\section{Hierarchical learning-based control}

\begin{figure*}[ht]
    \centering
    \vspace{0.1cm}
        \begin{tikzpicture}[node distance=1cm, auto, >=Stealth]

        \definecolor{h_blue}{RGB}{0, 0, 255}
        \definecolor{t_magenta}{RGB}{255, 0, 255}
        \definecolor{g_green}{RGB}{102, 255, 102}

        \node[rectangle, draw, rounded corners, align=center, fill=green!10, minimum width=3.5cm,
              minimum height=3.5cm, label={[green!60!black, above, yshift=-0cm]\textbf{Environment}},] (env) {};

        \node[rectangle, draw, rounded corners, align=center, fill=orange!30, minimum width=3.5cm, minimum height=2.6cm, left=1cm of env, label={[h_blue, above]\textbf{Herder $\B{i}$}},] (agent_i) {};
        
        \node[rectangle, draw, rounded corners, fill=yellow!40,
              align=center, minimum width=2.7cm, minimum height=1cm] at ($(agent_i) + (0, 0.6)$) (selection_i) 
              {Target-selection\\Policy};

        \node[rectangle, draw, rounded corners, fill=cyan!40,
              align=center, minimum width=1.6cm, minimum height=1cm,
              below=0.2cm of selection_i] (driving_i) {Driving\\Policy};

        \node[rectangle, draw, rounded corners, align=center, fill=orange!30, minimum width=3.5cm, minimum height=2.6cm, right=1cm of env, label={[h_blue, above]\textbf{Herder $\B{j}$}},] (agent_j) {};
        
        \node[rectangle, draw, rounded corners, fill=yellow!40,
              align=center, minimum width=2.7cm, minimum height=1cm] at ($(agent_j) + (0, 0.6)$) (selection_j) 
              {Target-selection\\Policy};

        \node[rectangle, draw, rounded corners, fill=cyan!40,
              align=center, minimum width=1.6cm, minimum height=1cm,
              below=0.2cm of selection_j] (driving_j) {Driving\\Policy};

  \node[circle, draw, fill=g_green, minimum size=1.1cm, inner sep=0pt, label={[below, font=\small]$\Omega_G$}] 
    at (env.center) {};

  \node[diamond, draw, fill=h_blue, minimum size=0.3cm, inner sep=0pt,
      label={[right, font=\small]$H_i$}]
    at ([shift={(0.5,-1.15)}]env.north west) (herder_i) {};
  \node[diamond, draw, fill=h_blue, minimum size=0.3cm, inner sep=0pt,
      label={[left, font=\small]$H_j$}]
    at ([shift={(3,-1.15)}]env.north west) (herder_j) {};
  
  \node[circle, draw, fill=t_magenta, minimum size=0.2cm, inner sep=0pt,
      label={[right, font=\small]$T_a$}]
    at ([shift={(-2.2,-0.4)}]env.north east) {};
  \node[circle, draw, fill=t_magenta, minimum size=0.2cm, inner sep=0pt,
      label={[right, font=\small]$T_b$}]
    at ([shift={(-2.7,-2.5)}]env.north east) {};
    \node[circle, draw, fill=t_magenta, minimum size=0.2cm, inner sep=0pt,
      label={[right, font=\small]$T_c$}]
    at ([shift={(-1.1,-2.2)}]env.north east) {};
    \node[circle, draw, fill=t_magenta, minimum size=0.2cm, inner sep=0pt,
      label={[right, font=\small]$T_d$}]
    at ([shift={(-1.7,-3)}]env.north east) {};
    \node[circle, draw, fill=t_magenta, minimum size=0.2cm, inner sep=0pt,
      label={[left, font=\small]$T_e$}]
    at ([shift={(-1.5,-1)}]env.north east) {};

        \draw[->] (selection_i.west) -| ($(selection_i.west) - (0.2,0)$)  |- node[right, yshift=0.3cm] {$\B{T}_i^*$} (driving_i.west);

        \draw[->] (driving_i.east)  node[above, xshift=1.4cm] {$\B{u}_i$} -| (herder_i.south);

        \draw[->] ($(env.south) - (1.25,0)$)  -| ++(0,-0.2) -| node[above, xshift=1cm] {$[\B{H}_j, \B{T}]$} (agent_i.south);

        \draw[->] (herder_i.west) -- node[above, xshift=0cm] {$\B{H}_i$} (selection_i.east);

        \draw[->] (selection_j.east) -| ($(selection_j.east) + (0.2,0)$)  |- node[left, yshift=0.3cm] {$\B{T}_j^*$} (driving_j.east);

        \draw[->] (driving_j.west)  node[above, xshift=-1.4cm] {$\B{u}_j$} -| (herder_j.south);

        \draw[->] ($(env.south) + (1.25,0)$)  -| ++(0,-0.2) -| node[above, xshift=-1cm] {$[\B{H}_i, \B{T}]$} (agent_j.south);

        \draw[->] (herder_j.east) -- node[above] {$\B{H}_j$} (selection_j.west);

    \end{tikzpicture}

    \vspace{0.3cm}
    
    \caption{Two-layer hierarchical feedback control scheme based on reinforcement learning (adapted from \cite{covone2025hierarchical}). Each herder $\mathbf{H}_{i,j}$ detects the other agents' positions and  selects the target $\mathbf{T}_{i,j}^*$ to control via the \textit{target-selection} policy, which is then driven according to the \textit{driving} policy, that outputs the velocity $\B{u}$ of the corresponding herder. 
    }
    \label{fig:control_architecture}
\end{figure*}
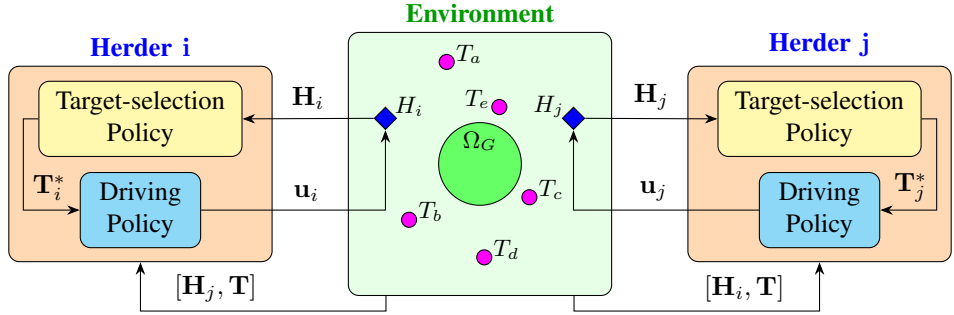

To address the problem complexity, we adopt a \textit{two-layer hierarchical control architecture}. As illustrated in Fig.~\ref{fig:control_architecture}, assuming each herder engages only with one target at a time \cite{aulettaHerdingStochasticAutonomous2022, ninoDeepAdaptiveIndirect2023, lamaShepherdingControlHerdability2024}, the overall task is decomposed into two interrelated subtasks: \textit{target selection} and \textit{target driving}. The high-level policy selects which target each herder should pursue, while the low-level policy drives the herder to interact with the selected target toward the goal region.
For the \textit{low-level control policy}, we train and compare both Deep Q-Network (DQN) and Proximal Policy Optimization (PPO) in a single herder-target scenario, allowing specialization in the driving subtask without multi-agent coordination complexity. We then fix the low-level controller and train the \textit{high-level decision-making policy} in a multi-agent environment. This layer requires coordinated target assignments, framing the problem as a Multi-Agent Reinforcement Learning (MARL) challenge. We adopt the \textit{centralized training with decentralized execution} (CTDE) paradigm \cite{gronauer_multi-agent_2022}, where agents train using shared information but act independently at execution time. Within this paradigm, we investigate multi-agent extensions of both DQN and PPO. Ultimately, the high-level policy determines the inputs to the low-level controller, which generates the herder's control actions.

\section{One Herder – One Target Scenario}
\label{sec:driving}
We begin by developing the low level policy that is used to complete the \textit{driving} subtask; to do so, we consider the case of a single herder and a single target  ($N=1$, $M=1$).
The goal of the herder is to learn to guide the target agent into a predefined goal region without prior knowledge of how its position influences target behaviour.

The learning agent receives as input its own coordinates and the coordinates of the target agent, and outputs a control action corresponding to the desired velocity vector for the herder, as shown in Fig. \ref{fig:control_architecture}.

Designing an effective reward function is crucial for ensuring that the policy learns purposeful behavior. While Reinforcement Learning can yield complex behaviors from simple reward signals (e.g., \cite{mnih2015human}), carefully shaped rewards can greatly improve sample efficiency, convergence stability, and potentially achieve analytical closed-loop performance guarantees \cite{heessEmergenceLocomotionBehaviours2017,10534075}. 

Hence, building upon results from the existing literature \cite{albi2016}, we design a reward function that captures four distinct objectives: (i) \textit{approaching} the target to enter its influence zone, (ii) 
\textit{steering} the target toward the goal, (iii) minimizing \textit{control effort}, and (iv) avoiding the \textit{herder entering} the goal. The resulting reward is defined as: 
\begin{equation}
    \begin{aligned}
    r_{\R{D}, k} =& - k_\R{a} \norm{\B{T}(t_k) - \B{H}(t_k)} + \\ & - k_\R{s} \left( \norm{\B{T}(t_k)} - \rho_G \right) \indicator{\domain \setminus \goalregion}{\T{}{t_k}} + \\ & -  k_\R{c} \norm{\B{u}(t_k)} - k_\R{h} \indicator{\goalregion}{\He{}{t_k}},
    \end{aligned}
    \label{eqn:reward_1H1T}
\end{equation} 
where $\mathds{1}_{A} : \domain \to \{0,1\}$ is the indicator function, defined by $\indicator{A}{\mathbf{x}}=1$ if $\mathbf{x} \in A$ and 0 otherwise for a given set $A \subseteq \domain$.

The gain values, were carefully chosen to reflect the relative importance of each behavioral objective, establishing a clear hierarchy: $k_\R{s}>0$ (goal guidance) was assigned the highest value to prioritize driving the target toward the goal. $k_\R{a}>0$ (target approach) was set to an intermediate value to accelerate early learning by encouraging the herder to enter the -- unknown -- target's zone of influence. $k_\R{c}>0$ (control efficiency) received the lowest weight, promoting minimal movement once the target is under control. 
This hierarchy ($k_\R{s} > k_\R{a} > k_\R{c}$) encourages progressive learning.
Finally, a sparse penalty term is applied via $k_\R{h}>0$ if the herder enters the goal region. This discourages it from disturbing other targets already inside the region in a multi-target setting.
The specific numerical values used in our simulations are provided in Appendix \ref{appendix:numerical}.

\subsection{Training the Deep Q-Network driving policy} 
To solve the driving sub-task using learning-based methods, we first train a Deep Q-Network (DQN) \cite{mnih2015human} that takes as input a four-dimensional observation vector $\mathbf{S}(t_k)=\left[\T{}{t_k}, \He{}{t_k}\right] \in \domain^2$, representing the absolute positions of the herder and target. This results in four neurons in the input layer of the neural network. The output corresponds to the $x$ and $y$ components of the herder’s discretized velocity vector. As DQN supports a continuous state space but requires a discrete action space, each velocity component is discretized into five bins uniformly: $\{-v_\R{H},-\frac{v_\R{H}}{2}, 0, \frac{v_\R{H}}{2}, v_\R{H}\}$. This choice yields $5^2$ possible discrete combinations, leading to $25$ neurons in the output layer.

Training is conducted over $E=5 \cdot 10^3$ episodes, with hyperparameters reported in Appendix \ref{appendix:numerical} and empirically tuned from initial values based on \cite{mnih2015human}.

\begin{figure}[t]
    \centering
    \vspace{0.3cm}
    \subfloat[]
    {
        \includegraphics{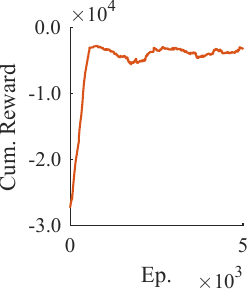}
        \label{fig:driving_dqn_rw}
    }
    \subfloat[]    
    {
        \includegraphics{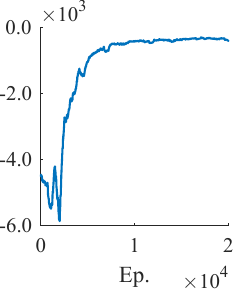}
        \label{fig:driving_ppo_rw}
    }
    \caption{Learning curves for the \textit{driving} policy during training. (a) DQN and (b) PPO agents' cumulative rewards per episode are shown. Only the first half of the training process is displayed to emphasize the learning phase. Curves are smoothed using a moving average over 100 samples for DQN and 1000 samples for PPO.}
    \label{fig:rewards_driving}
\end{figure}

Training results are shown in Fig.~\ref{fig:driving_dqn_rw}, where the cumulative reward per episode converges to a steady-state value within $5000$ episodes.

\subsection{Training the Proximal Policy Optimization driving policy}
As an alternative to DQN, we also implement PPO with continuous actions for the driving sub-task to enable smoother and more flexible herder motion.

The input to the policy network now consists of the absolute target position and the relative position between the herder and the target, both normalized by the initial domain radius $\domainradius$, resulting in a four-dimensional observation space $\mathbf{S}(t_k) = \left[ \T{}{t_k}, \T{}{t_k} - \He{}{t_k} \right]/\domainradius \in \domain^2$. While  observation space normalization with respect to $\rho_0$ is not strictly necessary, it helps with the stabilization and generalization of the Deep RL solution \cite{engstrom2019implementation}.

For any observation $\mathbf{S}(t_k)$ the actor outputs the means of the Gaussian policy $\pi(\mathbf{u}\mid\mathbf{S}(t_k))$, where the action $\mathbf{u}$ comprises two independent components along the $x$ and $y$ directions, each modeled by a separate univariate Gaussian distribution. The (log-)standard deviations of these distributions are not conditioned on the input but are learned as independent trainable parameters, as in \cite{schulmanTrustRegionPolicy2017}.
The critic receives the same observation to estimate a scalar state-value.
During training, actions are drawn from the Gaussian distribution to encourage exploration. In deployment, actions are deterministically selected as the means of the distributions.

Following \cite{andrychowiczWhatMattersOnPolicy2020, schulmanProximalPolicyOptimization2017}, we select hyperparameters for our PPO agent, which are reported in Appendix \ref{appendix:numerical}.

Training was carried out over $\numepisodes = 2 \cdot 10^4$ episodes. The learning curve in Fig.~\ref{fig:driving_ppo_rw} shows a sharp initial rise in cumulative reward, followed by convergence to a stable plateau.

\subsection{Validation}
\label{section:driving_validation}
Both \textit{driving} policies were evaluated over a batch of \(\numepisodes = 1000\) episodes with identical seeds to ensure consistent initial conditions. Performance was assessed using the metrics defined in Section~\ref{sec: metrics}.

\begin{figure}
    \centering
    \subfloat[]
{
    \includegraphics{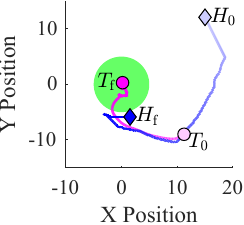}
    \label{fig:traj_DQN_driving}
}
\subfloat[]
{
    \includegraphics{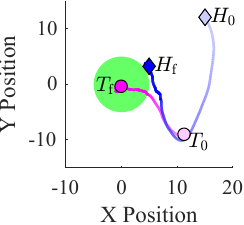}
    \label{fig:traj_PPO_driving}
}\\
\subfloat[]
{
    \includegraphics{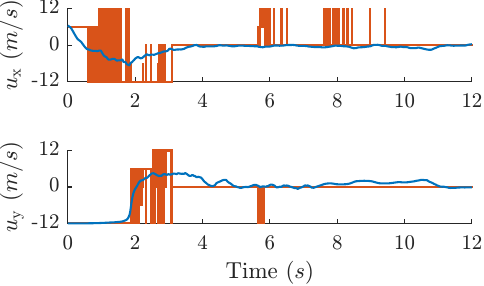}
    \label{fig:driving_task_example_velocities}
}

    \caption{Example of the learned \textit{driving} policy in a single-herder, single-target scenario for the (a) DQN and (b) PPO agents: the herder (blue diamond) approaches the target (magenta circle), guides it toward the goal region (green circle), and maintains containment. Color gradients represent the progression of positions over time, going from $t=0$ -- agents indicated as $\mathbf{H}_0=\mathbf{H}(0)$ and $\mathbf{T}_0=\mathbf{T}(0)$ -- to $t=t_f$ -- agents indicated as $\mathbf{H}_f=\mathbf{H}(t_f)$ and $\mathbf{T}_f=\mathbf{T}(t_f)$. In (c), the $x$ and $y$ velocity components are shown for both the DQN (orange) and PPO (blue) agents.}
    \label{fig:driving_task_example}
\end{figure}

Fig.~\ref{fig:driving_task_example} shows representative trajectories of the trained agents executing the \textit{driving} task. The learning-based herders consistently demonstrate  expected behavior as they first approach the target from behind, then guide it toward the goal region, and finally stabilize it with minimal movement, validating the design of our reward function and the effectiveness of our solution for the proposed setting. 

As shown in Fig.~\ref{fig:traj_PPO_driving}, PPO's continuous action space produces smoother and more natural herder trajectories. In contrast, Fig.\ref{fig:traj_DQN_driving} highlights DQN's efficiency during the containment phase, where it achieves target stabilization with fewer movements.
The main limitation of DQN stems from its discrete action space, as highlighted in Fig.~\ref{fig:driving_task_example_velocities}, where PPO varies control actions smoothly while DQN exhibits high-frequency switching during the gathering phase.

\begin{figure}
    \centering
    \subfloat[]
{
    \includegraphics{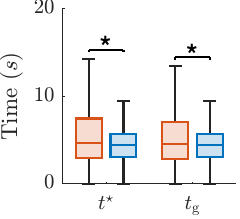}
    \label{fig:metrics_LL_time}
}
\subfloat[]
{
    \includegraphics{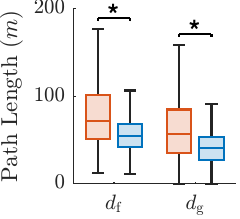}
    \label{fig:metrics_LL_space}
}

    \caption{Validation results for the DQN (orange) and PPO (blue) driving policies in the $N=1,\ M=1$ setting, over 1000 episodes with seeded initial conditions, showing (a) settling time $\settlingtime$ and gathering time $\gatheringtime$ in seconds and (b) final path length $\pathlengthfinal$ and gathering path length $\pathlengthgath$ in meters. 
    Box plots are shown for each metric. Mann-Whitney U test was performed on each metric pair yielding $p$-values always smaller than $0.001$.
}
    \label{fig:metrics_driving}
\end{figure}

Detailed performance metrics are shown in Fig.~\ref{fig:metrics_driving}, demonstrating our control strategy's effectiveness and consistency in the shepherding task with a single herder and single target. Both policies achieved 100\% success rates across all initial conditions; however, PPO consistently outperformed DQN on all evaluated metrics, as confirmed by Mann-Whitney U statistical tests. Although both agents exhibited similar completion times, PPO showed clear advantages in efficiency. This improvement is attributed to smoother trajectories from PPO's continuous action space, resulting in more effective guidance.

While DQN converges faster than PPO and is less sensitive to hyperparameter tuning \cite{pmlr-v48-duan16}, given PPO's superior performance, we adopt the PPO strategy as the baseline driving policy in subsequent sections.

\section{Multiple Herders – Multiple Targets Scenario}
\label{sec:target_selection}
We now address the \textit{target selection} task in the general case involving multiple herders and multiple targets ($\numherders > 1$, $\numtargets > 1$). In this setting, each herder must decide which target to engage with, taking into account the spatial distribution of all agents. Strategic coordination is essential to ensure an effective division of labor and to minimize redundant efforts among herders. In our multi-agent setting, each herder agent can sense the location of all the agents. However, during both training and deployment, herders are never given access to other herders' target selection choices. 

To enhance training efficiency and reduce computational complexity, we adopt the \textit{centralized training with decentralized execution} (CTDE) paradigm \cite{gronauer_multi-agent_2022}. In this framework, agents are trained using global information and shared learning mechanisms, while during deployment, each herder acts independently based solely on its local observations.

The goal is to learn a high-level policy that selects a target $\selectedtarget{i}{t_k}$ for each herder $i$ at every time step $t_k$, based on agent positions. 
As stated in the previous section, we use the PPO \textit{driving policy} as a fixed module, whose inputs are  determined by the high-level policy. 

During training, the target selection policy is queried only every $n_{\text{w}}$ time steps, allowing each herder to better observe the consequences of its selection and the resulting effects on the reward function, compared to a scenario where a new target is selected at every time step. During validation, this constraint is removed, allowing herders to switch targets freely throughout the episode.

Due to architectural constraints in the neural network input and output spaces, we consider, for the sake of simplicity, a specific instance of the problem with $\numherders=2$ herders and $\numtargets=5$ targets. Each herder receives as input its own position, the position of the other herder, and the positions of all targets. The policy outputs a discrete action corresponding to the index of the selected target. 

The reward function for the target selection task is defined as:
\begin{equation}
    \label{eqn:reward_L2}
    r_{\R{T}, k} = - k_\R{t} \sum_{i=1}^M \left(\norm{\T{i}{t_k}} - \rho_\R{G}\right) \indicator{\domain \setminus \goalregion}{\T{i}{t_k}},
\end{equation}
penalizing the distances of targets that remain outside the goal region $\Omega_\R{G}$ at time step \(t_k\). This encourages the herders to select and influence targets that contribute to faster convergence. 
The reward function is global, thus every herder receives the same reward signal, i.e. $r_{\R{T},k}^{(j)} = r_{\R{T},k} \,, \; \forall \, j=1,\dots,N$. %

\subsection{Training the Deep Q-Network target-selection policy} 
For the target selection sub-task, we define a Deep Q-Network (DQN) that receives as input the positions of two herders and five targets, resulting in $14$ neurons in the input layer, i.e., the observation vector of the $j$-th herder is $\mathbf{S}_j(t_k)=\left[\mathbf{H}_j(t_k), \mathbf{H}_{l \neq j}(t_k), \mathbf{T}(t_k) \right] \in \domain^{\numherders+\numtargets}$. The action space consists of the $\numtargets=5$ target indices, leading to five output neurons. 

We adopt the Deep Q-Network algorithm \cite{mnih2015human}, extended to multi-agent control via parameter sharing, following the approach in \cite{guptaCooperativeMultiagentControl2017}. 
In particular, we train a Deep Q-Network with a shared replay buffer that all herders contribute to.

Training is conducted over $\numepisodes=4 \cdot 10^5$ episodes, with  hyperparameters reported in Appendix \ref{appendix:numerical} and empirically fine-tuned from initial values based on the low-level training to cope with the non-stationarity of the multi-agent environment.

\begin{figure}[t]
    \centering
    \includegraphics{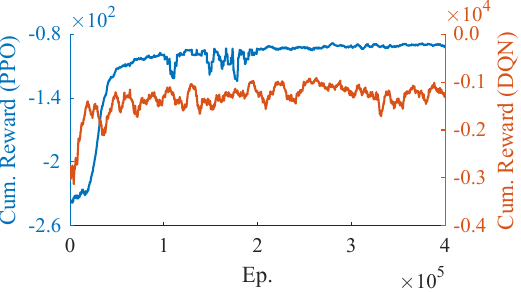}
    \caption{Learning curves for the \textit{target-selection} policy during training. DQN (orange) and PPO (blue) agents' cumulative rewards per episode are shown. Curves are smoothed using a moving average over 5000 samples. }
    \label{fig:rewards_selection}
\end{figure}

Fig.~\ref{fig:rewards_selection} shows that the cumulative reward per episode converges to a steady value during training, indicating that the model learns an effective and consistent decision-making policy to solve the proposed sub-task.

\subsection{Training the Proximal Policy Optimization target-selection policy}
\label{subsec:target_selection_ppo}

We also implement Multi-Agent Proximal Policy Optimization (MAPPO) \cite{yuSurprisingEffectivenessPPO2022} using an Actor-Critic architecture, where both the actor and critic networks share the same structural design. The input to both networks consists of the same $2(\numherders + \numtargets)$ features used in the DQN implementation, normalized to the environment dimensions to improve numerical stability.

The actor network outputs a probability distribution over the $\numtargets$ selectable targets using softmax activation, enabling stochastic action selection during training. The critic network produces a scalar value estimate for the given state. 
During deployment, softmax activation is bypassed and agents deterministically select targets corresponding to maximum-valued actor outputs, ensuring consistent behavior.

Training was carried out over $\numepisodes = 2 \cdot 10^5$ episodes. The  learning curve  in Fig.~\ref{fig:rewards_selection} shows stable and progressive improvement in agent performance throughout training. Hyperparameters, reported in Appendix \ref{appendix:numerical}, were initially based on single-agent PPO for the driving task, then tuned for the multi-agent shepherding scenario's increased complexity.

Following recommendations from \cite{yuSurprisingEffectivenessPPO2022}, training used no minibatching, as this yields better multi-agent performance. Additionally, we disabled entropy regularization, since policy stochasticity from softmax outputs provided sufficient exploration without explicit entropy bonuses.

\subsection{Validation}
\begin{figure}
    \centering
    \subfloat[]
{
    \includegraphics{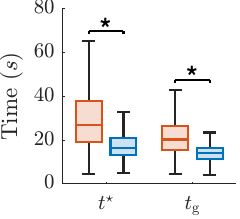}
    \label{fig:metrics_HL_time}
}
\subfloat[]
{
    \includegraphics{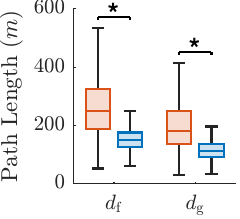}
    \label{fig:metrics_HL_space}
}\\

\subfloat[]
{
    \includegraphics{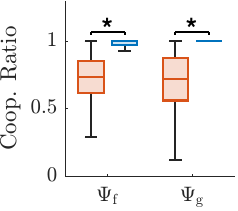}
    \label{fig:metrics_HL_coop}
}

    \caption{Validation results for the DQN (orange) and MAPPO (blue) target selection policies in the $N=2,\ M=5$ setting, over 1000 episodes with seeded initial conditions, showing (a) settling time $\settlingtime$ and gathering time $\gatheringtime$ in seconds, (b) final path length $\pathlengthfinal$ and gathering path length $\pathlengthgath$ in meters and (c) average cooperation at final time $\coopratiofinal$ and gathering time $\coopratiogath$. 
    Box plots are shown for each metric. Mann-Whitney U test was performed on each metric pair yielding $p$-values always smaller than $0.001$.}
    \label{fig:metrics_selection}
\end{figure}

We evaluate both the DQN and MAPPO policies over \(\numepisodes = 1000\) test episodes, using seeded initial conditions uniformly sampled from $\initialdomain$.
As shown in Fig.~\ref{fig:metrics_selection}, both RL strategies successfully learn effective control behaviors. In particular, Fig.~\ref{fig:metrics_HL_coop} reveals that agents spontaneously develop cooperative strategies by primarily selecting different targets with minimal overlap, improving spatial coverage and accelerating task completion. Remarkably, this division of labor is not encoded in the reward function but emerges naturally from the learning process. The strong coordination observed,  particularly until the gathering time, demonstrates the ability of reinforcement learning agents to develop complementary roles without explicit communication or predefined rules. 

\begin{figure}
    \centering
    \subfloat[]
{
    \includegraphics{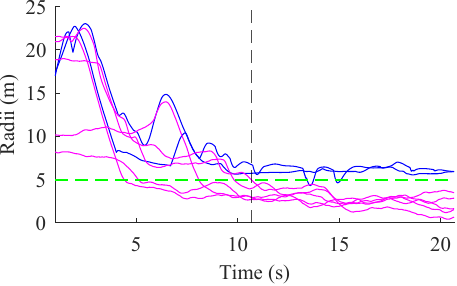}
    \label{fig:selection_radii}
}\\
\subfloat[]
{
    \includegraphics{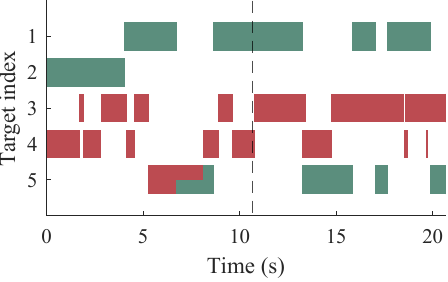}
    \label{fig:selection_choices}
}

    \caption{Validation example of the MAPPO solution in the scenario with $\numherders = 2$ herders and $\numtargets = 5$ targets.
    (a) Temporal evolution of the distances of herders (blue lines) and targets (magenta lines) from the center of the goal region. The green dashed line indicates the goal region threshold \(\goalradius = 5\). The plot illustrates how the herders successfully guide and contain all targets within the goal region.
    (b) Target selection over time for the two herders, with selections shown in green and red respectively. The plot highlights the emergence of cooperative behavior, as the two agents predominantly select different targets with minimal overlap, enabling efficient and coordinated shepherding.
    In both panels, the vertical dashed line indicates the gathering time.}

    \label{fig:example_selection}
\end{figure}

Fig.~\ref{fig:example_selection} shows a representative episode where two herders, controlled by the MAPPO policy, successfully coordinate to guide and contain five target agents into the goal region. The evolution of target distances from the goal center is shown in Fig.~\ref{fig:selection_radii}, where all radii fall below the threshold \(\goalradius\), confirming both effective gathering and stable containment over time.  Fig.~\ref{fig:selection_choices} provides insights into the herders’ decision-making dynamics, showing herders selecting different targets to cooperate effectively and enhance efficiency. We also observe that although herders query the high-level policy at every time step, they do not switch targets at each step, creating effective time-scale separation that enables efficient task execution.

MAPPO consistently outperforms DQN across key performance metrics, including settling time, control efficiency, and average cooperation. These improvements appear not only in mean values, but also in lower variability across episodes, indicating greater robustness and consistency in MAPPO's decision-making. Consequently, we adopt the trained MAPPO policy for high-level decision-making.

\subsection{Theoretical Challenges and Validation Strategy}
\label{sec:theoretical_considerations}

Formal convergence analysis for the proposed hierarchical multi-agent reinforcement learning (MARL) architecture is intractable due to three main factors. 

First, the system operates in an inherently non-stationary environment: every policy update by one agent perturbs the transition kernel perceived by the others \cite{gronauer_multi-agent_2022}. While single-agent RL algorithms can provide convergence guarantees under restrictive assumptions such as stationary Markovian dynamics \cite{sutton2018reinforcement, bertsekas2019reinforcement}, these assumptions do not hold in MARL, making theoretical guarantees a central open challenge \cite{gronauer_multi-agent_2022}.

Second, the architecture exhibits hybrid discrete–continuous dynamics, as each herder must simultaneously select discrete target assignments and generate continuous control actions. While bounded single-herder dynamics have been analyzed in~\cite{licitraSingleAgentIndirectherding2018} with a rule-based target selection policy, extending such analysis to the multi-herder RL setting with coupled switching and continuous evolution quickly becomes analytically intractable.

Third, our sequential training strategy first optimizes the driving policy on single herder–target pairs, then fixes it while training the target-selection policy. This approach mitigates the prohibitive sample complexity of fully joint training \cite{nguyen2019deep}, but it inherently introduces a suboptimality gap and further complicates any formal analysis of optimality or stability.

Given these challenges, we rely on a comprehensive empirical validation strategy that demonstrates the stability and effectiveness of the learned policies through (i) multi-scenario simulation studies, (ii) systematic benchmarking against state-of-the-art methods, and (iii) real-world experiments on the Robotarium platform. 

\section{Simulation results and analysis}

As discussed in Section \ref{sec:introduction}, limited research has addressed shepherding with non-cohesive targets. We benchmark against two representative approaches: a heuristic strategy for non-cohesive targets \cite{aulettaHerdingStochasticAutonomous2022} and a model-based approach designed for cohesive targets \cite{piersonControllingNoncooperativeHerds2018}. Then we study scalability to larger settings and extend the approach to solve a tracking scenario.

\subsection{Benchmarking against state-of-the-art approaches}
\paragraph{Heuristic approach for non-cohesive targets}
Auletta et al. \cite{aulettaHerdingStochasticAutonomous2022} decompose the task into target selection and driving components, similar to our architecture. We focus on their dynamic Peer-to-Peer (P2P) strategy, where herders dynamically partition the plane and select the furthest target in their sector, explicitly enforcing cooperation. For driving, each herder positions itself behind the selected target at a fixed distance encoded in the model. Unlike \cite{aulettaHerdingStochasticAutonomous2022}, we assume no explicit knowledge of system dynamics.

\begin{figure}
    \centering
    \subfloat[]
{
    \includegraphics{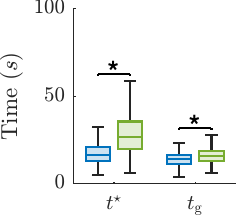}
    \label{fig:metrics_time_comparison}
}
    \subfloat[]
{
    \includegraphics{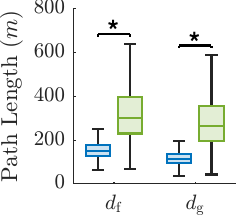}
    \label{fig:metrics_space_comparison}
}\\

    \subfloat[]
{
    \includegraphics{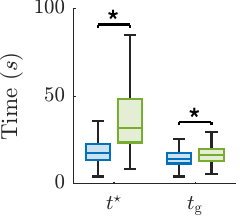}
    \label{fig:metrics_time_robustness}
}
    \subfloat[]
{
    \includegraphics{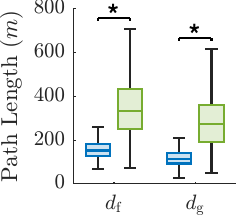}
    \label{fig:metrics_space_robustness}
}

    \caption{
    Validation results comparing the learning-based strategy (blue) with the heuristic approach (green) from \cite{aulettaHerdingStochasticAutonomous2022}. Results are averaged over 1000 episodes with seeded initial conditions. Box plots show: gathering time $\gatheringtime$ and settling time $\settlingtime$ in seconds (panels a, c), and path length at gathering time $\pathlengthgath$ and final time $\pathlengthfinal$ in meters (panels b, d). Panels (a-b) correspond to the nominal parameter setting, while panels (c-d) report a robustness analysis with target model parameters varied by $\pm20\%$ around their nominal values. A Mann-Whitney U test was performed on each metric pair, yielding p-values smaller than 0.001, indicating statistically significant differences (*).}
    \label{fig:metrics_benchmark_comparison}
\end{figure}

For each scenario, we evaluate 1000 episodes with identical seeded initial conditions using 2 herders and 5 targets. Figures \ref{fig:metrics_time_comparison}-\ref{fig:metrics_space_comparison} show that both policies achieve 100\% success rates, but our learning-based strategy outperforms the heuristic baseline in completion time and efficiency.

While the heuristic strategy remains effective during gathering, it struggles to maintain targets within the goal region, leading to completion times significantly longer than gathering times. In contrast, our RL policy employs active containment preventing target escape, resulting in shorter overall completion times. Regarding efficiency, the heuristic strategy's requirement to always select the farthest target causes herders to oscillate between distant targets, while our RL policy balances target distance with herder proximity through optimization.

\paragraph{Model-based approach for cohesive targets}
The approach by Pierson and Schwager \cite{piersonControllingNoncooperativeHerds2018} relies on cohesion forces among targets and controls the herd using formation-based methods. As expected, this approach fails in our non-cohesive setup, achieving only 8.7\% success rate. 

For a fair comparison, we also evaluate both approaches under cohesive conditions using the dynamics from Vaughan et al. \cite{vaughan2000experiments}. Targets are initialized within a small neighborhood (radius 2) around their center of mass. Under these conditions, \cite{piersonControllingNoncooperativeHerds2018} achieves 96.8\% success rate with gathering time 28.01 ± 23.67s, while our strategy attains 99.8\% success rate with significantly lower gathering time of 11.85 ± 6.87s.

\paragraph{Robustness analysis}
We assess robustness by perturbing target dynamics parameters (\(\diffusioncoeff\), \(\zeta\), \(\longrangecoeff\)) sampled from Gaussian distributions with 20\% standard deviation around nominal values. We test this only for our approach versus the non-cohesive heuristic baseline \cite{aulettaHerdingStochasticAutonomous2022}, since the cohesive approach \cite{piersonControllingNoncooperativeHerds2018} already fails under nominal conditions. Results in Figures \ref{fig:metrics_time_robustness}-\ref{fig:metrics_space_robustness} confirm that both strategies show strong robustness to parametric uncertainties, but our RL strategy maintains superior performance with 99.8\% success rate versus 98.7\% for the heuristic approach.

These results demonstrate that: (i) strategies designed for cohesive targets fail when this assumption is removed, and (ii) (ii) our learned policy, trained without cohesion assumptions, performs effectively across both cohesive and non-cohesive scenarios, not only solving the more challenging non-cohesive case but also outperforming specialized heuristics even in their intended cohesive setting. This versatility demonstrates the robustness and generalizability of learning-based approaches over model-based or heuristic strategies.

\subsection{Scalability to large-scale settings} 
We extend our strategy to address scalability challenges from training constraints. The number of agents in training defines the neural network architecture, limiting each herder to observing and acting upon a fixed number of agents. To overcome this limitation, we adopt limited topological sensing, where each herder observes only a number of its nearest neighbors, set to be equal in number to those considered during training; preserving, therefore, compatibility with the fixed dimensions of the network. This approach allows policies trained in smaller environments to scale to larger systems, mitigating the curse of dimensionality.

We demonstrate this approach with 5 herders corralling 50 targets. Each herder receives the closest herder position and five nearest target positions as input, enabling direct application of policies trained in Sections~\ref{sec:driving}–\ref{sec:target_selection} without retraining.

\begin{figure}
    \centering
    \includegraphics{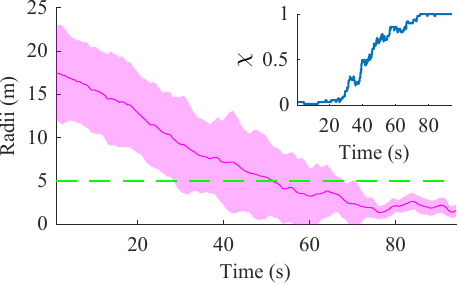}
    \caption{Extension of the trained policy in a large-scale setting with $\numherders=5$ herders and $\numtargets=50$ targets via topological sensing. The evolution of the mean target distance from the goal (magenta solid line) along with the standard deviation (magenta shaded area) is shown relative to the goal region radius (green dashed line). The inset displays the evolution of the fraction of captured targets $\chi$. The mean target distance from the goal falls below the goal region radius, indicating successful task completion, as confirmed by $\chi$ reaching 1.}
    \label{fig:scalability_example}
\end{figure}

Fig.~\ref{fig:scalability_example} shows successful task completion: average target distance from the goal drops below the goal region radius, and the fraction of contained targets $\chi$ reaches 1, thus indicating successful completion of the task.

\subsection{Extension to a tracking scenario}

We demonstrate our solution's flexibility by addressing the problem of guiding the targets towards time-varying goal regions. While standard shepherding tasks assume fixed goal regions, some studies \cite{van2024reactive} explore steering agents along predefined safe paths, effectively formulating multi-agent tracking control problems. We allow the goal region center $\goalcenter{t}$ to evolve over time rather than remaining fixed. The goal region center represents the nominal target trajectory, while the goal radius $\goalradius$
represents the allowable deviation from this path, i.e., the safe boundary layer width.

Through reference frame transformation, we avoid retraining the policies from Sections \ref{sec:driving}--\ref{sec:target_selection}.  We shift the observation vectors relative to $\goalcenter{t}$ and feed them to the already trained neural networks. First, we constrain $\goalregion(t) \in \initialdomain$ for all $t$, ensuring that the goal region remains within the domain encountered during training. Second, we assume a time-scale separation between the agents' dynamics and those of the moving goal region, setting their speed ratio to $1:50$. This ratio provides a reasonable balance between agent responsiveness and tracking performance in our experiments.

\begin{figure}
    \centering
    {\includegraphics{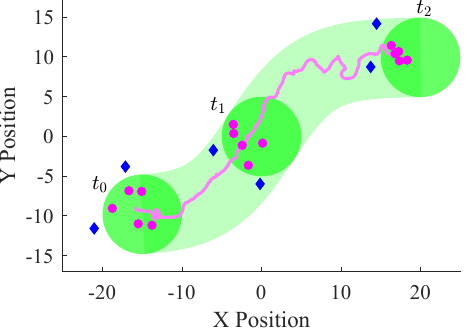}}
    \caption{Validation example of the proposed strategy in a tracking scenario with $\numherders=2$ herders and $\numtargets=5$ targets. The goal region (green area) moves along a sigmoid-like trajectory (shaded green path). The mean trajectory of the targets is shown as a magenta solid line, consistently remaining within the desired safe path. Three snapshots are highlighted at the initial ($t_0$), middle ($t_1$), and final ($t_3$) times of the simulation, demonstrating that the herders (blue diamonds) successfully contain the targets (magenta dots) within the time-varying goal region throughout the task.}
    \label{fig:tracking_example}
\end{figure}

Without loss of generality, we consider a sigmoid-like trajectory for  $\goalcenter{t}$ that begins in the bottom-left corner of $\initialdomain$ and progresses toward the upper-right. Fig.~\ref{fig:tracking_example} shows the mean trajectory of the targets, which remains consistently within the safe path, demonstrating the herders' ability to maintain containment and efficiently solve the tracking problem.

\section{Experimental validation}
To demonstrate our strategy's effectiveness in real robotics settings, we complement numerical simulations with experiments on real robots using the online Robotarium platform \cite{wilson_robotarium_2020}.

\begin{figure}
    \centering
    \subfloat[]
{
    \includegraphics[width=0.45\linewidth]{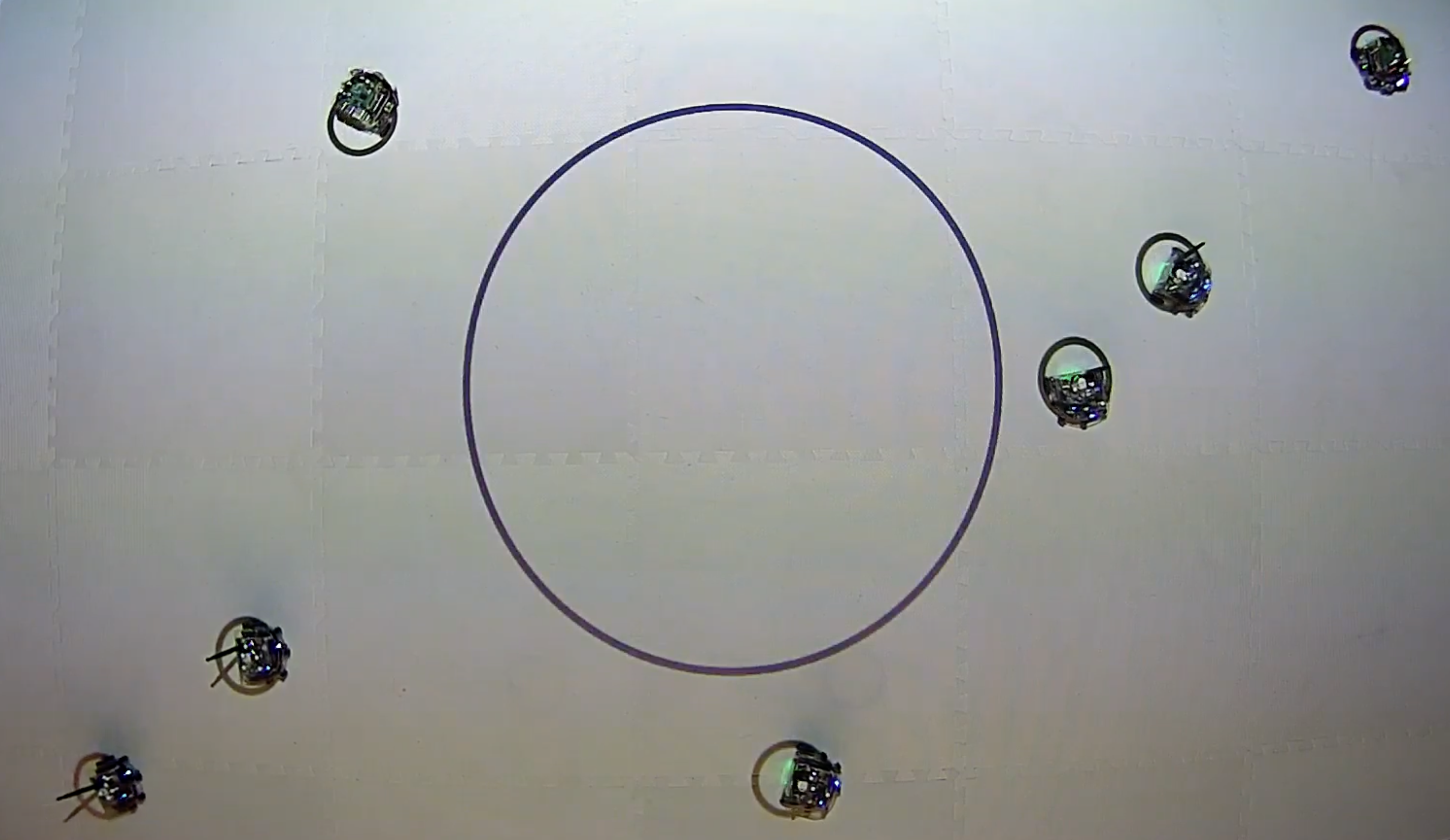}
    \label{fig:robot_initial}
}
    \subfloat[]
{
    \includegraphics[width=0.45\linewidth]{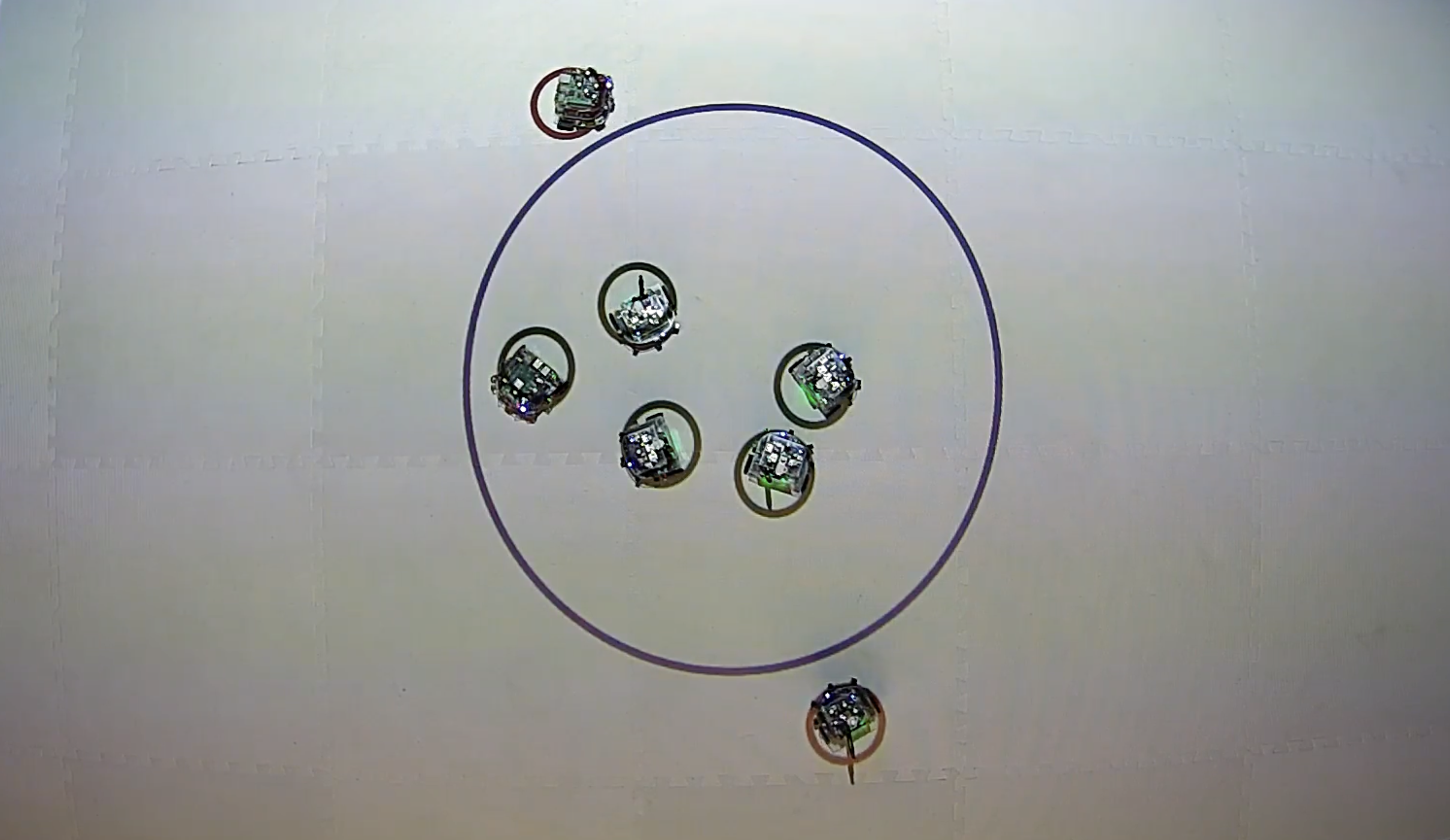}
    \label{fig:robot_final}
}\\

    \subfloat[]
{
    \includegraphics{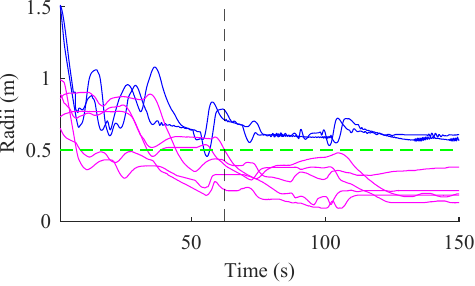}
    \label{fig:robotarium_radii}
}\\
    \subfloat[]
{
    \includegraphics{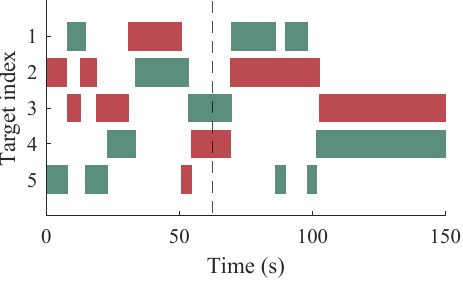}
    \label{fig:robotarium_selection}
}

    \caption{
Experimental validation of the RL-based shepherding strategy on the Robotarium platform. (a) Initial configuration with two herder robots placed at the top-right and bottom-left corners and five target robots positioned centrally outside the goal region. (b) Final configuration showing all targets successfully contained within the goal region. (c) Time evolution of the radial distances of the targets from the goal center. The green dashed line denotes the goal radius ($\goalradius = 0.5\, m$), and the vertical dashed line indicates the settling time ($\settlingtime = 62.3\, s$). (d) Target selection over time for the two herders, showing effective cooperation without selecting the same target simultaneously (each color corresponds to one herder).
}
    \label{fig:robotarium_example}
\end{figure}

Robotarium is a remotely accessible research facility with GRITSBot robots, enabling rapid deployment and testing of custom control algorithms in multi-robot scenarios. In our experiments, we consider a setup with $\numtargets=5$ target robots and $\numherders=2$ herder robots. Due to
the arena's limited workspace ($3.2\, \text{m} \times 2\, \text{m}$) and the safety protocols preventing collisions (each robot is 11 cm diameter), we place herders at top-right and bottom-left corners while targets are positioned centrally outside the goal region, whose radius is set to $\goalradius = 0.5\, m$, as shown in Fig.~\ref{fig:robot_initial}.

To ensure feasibility given the hardware constraints of the GRITSBot robots, which have a maximum linear speed of 20 cm/s and a maximum rotational speed of approximately 3.6 rad/s, we scale the target dynamics and the herders’ observations accordingly, as detailed in Appendix \ref{appendix:numerical}.

Fig.~\ref{fig:robotarium_example} reports experimental results. Panel \ref{fig:robotarium_radii} shows herder robots steering targets into the goal region within $\settlingtime = 62.3 \,s$. Once the targets are inside the goal region, they are effectively contained until the end of the experiment, reaching the final configuration shown in Fig.~\ref{fig:robot_final}.
Fig.~\ref{fig:robotarium_selection} shows herder target selection at each time step, demonstrating effective cooperation with no simultaneous target selection.

The experiment demonstrates our RL-based policy's robustness, adaptability, and real-world feasibility. Despite training in simplified simulation with single-integrator herders and second-order targets, the policy successfully transfers to real robots with unmodeled unicycle dynamics, actuator limitations, and sensing uncertainties. This confirms our approach generalizes beyond training conditions and suits practical multi-robot shepherding deployment, even under physical and operational constraints.

\section{Conclusions}
We presented a fully decentralized, hierarchical reinforcement learning framework to solve the multi-agent shepherding problem without cohesion assumptions. An RL-based low-level \emph{driving} controller is combined with a MARL-based high-level \emph{target-selection} policy using both DQN and PPO. Without requiring explicit inter-herder communication or prior knowledge of target dynamics, the method consistently gathers, contains, and tracks non-cohesive stochastic targets, with spontaneous cooperation emerging among herders. 

Our approach outperforms state-of-the-art solutions and demonstrates flexibility with time-varying goal regions.
Large-scale simulations show the policy generalizes to significantly larger target groups, even with limited topological sensing. Robotarium experiments confirm seamless transfer to real differential-drive robots despite sensing noise and actuation constraints, highlighting hierarchical deep reinforcement learning's practical value for distributed multi-robot control.

While formal theoretical analysis remains challenging due to the non-stationary multi-agent environment and hierarchical architecture, our comprehensive empirical validation provides strong evidence of convergence and stability. The emergent cooperative behaviors observed suggest underlying coordination mechanisms that warrant future theoretical investigation.

Future work could enhance scalability through strategies tailored for large-scale multi-agent systems and higher-dimensional spaces. Incorporating restricted sensing, environmental obstacles, and adversarial targets would increase realism. Most importantly, establishing theoretical frameworks bounding the performance gap between learned policies and optimal solutions would provide essential safety and stability guarantees for real-world deployment. A possible way forward might be to move from agent-based descriptions of the problem to continuum descriptions as recently proposed in \cite{lamaInterpretableContinuumFramework2025},\cite{dilorenzo2024continuification},\cite{yang2018mean}.

\section*{Acknowledgments}
The authors acknowledge support from the Italian Ministry of University and Research (MUR) under project PRIN 2022 ``Machine-learning based control of complex multi-agent systems for search and rescue operations in natural disasters (MENTOR).''
The authors thank the Georgia Institute of Technology for providing access to the Robotarium platform for experimental validation.
\bibliographystyle{ieeetr}
\bibliography{refs.bib}

\appendix
\renewcommand{\thetable}{A\arabic{table}}
\setcounter{table}{0} 
\label{appendix:numerical}
This appendix summarizes the key numerical settings used in the study.

\paragraph{Model and learning parameters}
Table \ref{tab: parameters} reports the physical constants defining herder and target dynamics (Section \ref{sec: mathematization}). Table \ref{tab:hyperparameters_ppo_mappo} lists hyperparameters for the reinforcement learning algorithms (Sections \ref{sec:driving}–\ref{sec:target_selection}), with reward gains in Table \ref{tab:reward_gains}.

\paragraph{Training protocol}
Training uses enlarged simulation steps (\(\samplingtimetrain\)) while validation employs nominal steps (\(\samplingtime\)). Coarser steps accelerate convergence by providing larger state transitions; for numerical stability, we set \(\shortrangecoeff = 0\) during training. Driving episodes terminate when targets remain in the goal region for \(\containmenttime = 10\mathrm{s}\) or \(\maxtime = 60\mathrm{s}\) is reached. Target-selection extends the horizon to \(\maxtime = 150\mathrm{s}\).

\paragraph{Network architectures}
Deep Q-Networks use ReLU activation in two hidden layers with linear output layers. Driving networks have 256 and 128 units; target-selection networks use 512 and 256 units. PPO actor-critic networks employ five hidden layers of 64 ReLU units with identical structures but separate parameters. Actors use final \(\tanh\) layers for bivariate Gaussian distributions; critics use linear scalar outputs. MAPPO uses two hidden layers with 256 and 128 ReLU units.

\paragraph{Robotarium scaling}
For GRITSBot compatibility, we scale target dynamics to $(\zeta, \diffusioncoeff, \longrangecoeff) = (60, 0, 0.1)$ and apply 1:10 spatial scaling. Robot positions from $[-1.6\, m, 1.6\, m] \times [-1\, m, 1\, m]$ are scaled to $[-16\, m, 16\, m] \times [-10\, m, 10\, m]$ for neural network input.

\paragraph{Implementation}
Simulations use Python with PyTorch neural networks and Gymnasium RL environments. Supplementary videos are available at \href{https://shorturl.at/IBjKK}{https://shorturl.at/IBjKK}. %
\begin{table}[h]
    \centering
    \caption{Model parameters for herders, targets, and environment used in the simulation study}
    \label{tab: parameters}
    \begin{tabular}{cc}
        \toprule
        Parameter & Value \\
        \midrule
        $\diffusioncoeff$ & 3\\
        $\longrangecoeff$ & 40\\
        $\shortrangecoeff$ & 40\\
        $\zeta$ & 4 \\
        $r_\mathrm{c}$ & 0.1 \\
        $v_{\mathrm{H}}$ & 12\\
        $\domainradius$ & 25 \\
        $\goalradius$ & 5 \\
        $\samplingtime$ & 0.01\\
        $\samplingtimetrain$ & 0.05\\
        \bottomrule
        
    \end{tabular}
\end{table}

\begin{table}
    \centering
    \caption{Hyperparameters of the RL training algorithms (values for the \textit{target selection} policies are indicated in parentheses only when different from the \textit{driving} ones). The parameters names refer to the nomenclature found in \cite{schulmanProximalPolicyOptimization2017} and \cite{mnih2015human}.}
    \begin{tabular}{ll}
        \toprule
        Hyperparameter & Value \\
        \midrule
        \multicolumn{2}{c}{\textbf{DQN}}\\
        \midrule
        Adam stepsize & 5e-5 (1e-4)\\
        Discount rate & 0.99 \\
        Initial exploration rate $\varepsilon$ & 1 \\
        Minimum exploration rate $\varepsilon$ & 0.05 \\
        Exploration rate decay & 1e-3 (5e-5)\\
        Target network update frequency & 1e4 (1e3)\\
        Minibatch size & 64\\
        \midrule
        \multicolumn{2}{c}{\textbf{PPO}}\\
        \midrule
        Adam stepsize & 5e-4 \\
        Discount rate & 0.98 \\
        GAE parameter & 0.95\\
        Clipping parameter & 0.2 \\
        VF coeff. & 0.5\\
        Entropy coeff. & 0.1 (0) \\
        Number of epochs & 10\\
        Horizon & 4096 (32)\\
        Minibatch size  & 128 (1024)\\
        Number of actors & 8 (32) \\
        \bottomrule
        
    \end{tabular}
    
    \label{tab:hyperparameters_ppo_mappo}
\end{table}

\begin{table}
    \centering
    \caption{Reward gains used by each reinforcement-learning algorithm}
    \begin{tabular}{lll}
        \toprule
        Hyperparameter & DQN & PPO \\
        \midrule
        $k_\R{a}$ & 0.5 & 0.05 \\
        $k_\R{s}$ & 1 & 0.1 \\
        $k_\R{c}$ & 0.1 & 0.001 \\        
        $k_\R{h}$ & 5 & 5 \\        
        $k_\R{t}$ & 1 & 0.01 \\        
        \bottomrule
        
    \end{tabular}
    
    \label{tab:reward_gains}
\end{table}
\end{document}